# Phonon propagation scale and nanoscale order in vitreous silica from Raman spectroscopy

Vitaly I. Korepanov

Intitute of Microelectronics Technology and High Purity Materials, RAS

For nanoscale systems such as nanoparticles and 3D-bonded networks, the range of spatial coherence is well reflected in the Raman spectral pattern. For confined, or localized, phonons, the range of *q*-points contributing to the spectrum depends on the phonon confinement length, which makes possible to derive the size information from the spectra. In this work, Raman spectrum of vitreous silica is described as localized phonons of $SiO_2$ network. The convergence of the spectral pattern with the confinement size is studied. It is shown that the phonon propagation scale in vitreous silica is within the 0.5-2 nm range.

## Introduction

Glasses possess a certain degree of local order as evidenced, in particular, by X-ray diffraction[1,2] and computational studies[3–7]. How to quantify the local order, and how to link the experimental data with understanding of the structure are disputable questions of high interest.

It is argued in the literature that Boson peak in Raman spectra can reflect the nanometer-scale local order[7–11]. This broad spectral feature if observed in the low-wavenumber range (LWR) (60-100 cm$^{-1}$). The bands in Raman spectra of such systems as fused silica and water represent the phonon modes localized by the disorder[11–13]. In glasses, the early estimation of the localization length was calculated to be in the order of few interatomic spacings[10].

The coherence range of the continuous network can be estimated from the propagation of phonon-like modes[14]. Raman spectral pattern for nanoscale systems is a reflection of this coherence[8,15]. For the localized (confined) phonons, the Raman spectrum is not limited to the near Brillouin zone (BZ) center modes. The contribution of different *k*-points depends on the size, and can be found from Fourier decomposition of the confined phonons into the original bulk wavefunctions[15]:

$$I(\omega) \cong \iiint \frac{\Gamma_0(\sigma) * |C(\boldsymbol{q_0},\boldsymbol{q})|^2}{(\omega - \omega(\boldsymbol{q}))^2 + (\Gamma_0(\sigma)/2)^2} d^3\boldsymbol{q} \quad , \quad (1)$$

where $\omega$ is the wavenumber, $\sigma$ is the confinement size, $\Gamma_0$ is the natural linewidth, $C(\boldsymbol{q_0},\boldsymbol{q})$ is the Fourier coefficient for a given confinement shape and size at a given wave vector $\boldsymbol{q}$. The scattering intensity at $\boldsymbol{q_0}$ depends on the phonon propagation direction, and can be calculated from Raman tensor[16]. This gives additional intensity factor $A_i(q, \varphi, \vartheta)$ for scattering angles $\varphi$ and $\vartheta$ [17]. The intensities should be integrated (numerically) over the scattering directions in polar coordinates and summed up over band index *i*:

$$I(\omega) \cong \sum_i \iiint \frac{A_i(\boldsymbol{q_0},\varphi,\theta) * \Gamma_{0,i} * |C(\boldsymbol{q_0},\boldsymbol{q})|^2 * q^2}{(\omega - \omega_i(\boldsymbol{q}))^2 + (\Gamma_{0,i}(\sigma)/2)^2} dq\, d\varphi\, d\theta \quad , \quad (2)$$

Within such approach, the acoustic phonons should have no contribution, because the intensity factors are zero. However, in disordered systems, the Raman intensity for acoustic phonon range is described a different way. It was shown that it has additional wavenumber dependence, referred to as light-to-vibration coupling coefficient $C_{ac}(\omega)$ [18]. Experimental studies showed that it has a linear wavenumber dependence $C_{ac}(\omega) \sim \omega$ [19,20].

The phonon confinement model formulated in this way was recently applied to hydrogen-bond network of water, it was shown that experimental Raman spectra of liquid water contain highly important information on the size of ice-like structure fragments[13]. The equation (2) links the Raman spectral pattern with the coherence range of the confined phonons. The present study aims at interpretation of the Raman spectra of silica in terms of phonons propagation scale of the $SiO_2$ network.

## Modeling and experimental details

Based on the similarities between cristobalite and fused silica glass in terms of structure and Raman spectra[3,7,21,22], cristobalite is taken as the structure motif at the nanometer scale for fused silica. The phonon dispersion $\omega(\boldsymbol{q})$ was calculated with the Quantum Espresso package[23,24]. The PBEsol functional was taken with high-throughput ultrasoft pseudopotentials[25] with cut-off for wavefunctions/charge density of 56/320 Ry. The 12-atom unit cell was taken. The geometry and cell parameters were optimized with 8x8x6 *k*-point grid; phonon dispersion was calculated with the 8x8x6 grid. Raman tensor was calculated with LDA (PZ) functional with norm-conserving pseudopotentials[26] with the cut-off of 112/448 Ry. The intensity factors $A_i(q, \varphi, \vartheta)$ were calculated from Raman tensor for the vibrational forms for each $\varphi$ and $\vartheta$. For each band, the intensity factors were scaled according to the experimental spectra of bulk cristobalite. Scaling of the phonon energies was done to match the computed Raman spectral pattern of cristobalite with the experimental data by multiplying the whole dispersion function of the given band $\omega_i(\boldsymbol{q})$ by the scaling factor[27]. The scaling factors were in the range (0.96…1.06). Calculated phonon dispersion is in good agreement with recent ab initio studies[7], where single scaling factor (1.039) was applied.

For the form-factor calculation, spherical confinement was assumed; the Fourier coefficients $C(\boldsymbol{q_0},\boldsymbol{q})$ for this case have an analytic form of Bessel-like function[27]. The integration in eq. (2) is done within [0,2] range of the reduced wavevector (i.e. not limited to the 1$^{st}$ Brillouin zone[28]).

The experimental spectra discussed here were corrected by the frequency factor $(\omega-\omega_0)^3$ and Boltzmann distribution factor[29]. For the correct comparison with the calculated spectra, the latter should also be divided by $\omega$ according to the Placzek's expression[30].

The Raman spectra of the localized phonons were calculated with eq. 2. For the natural linewidth, the inverse size dependence was assumed $\Gamma_i(\sigma)=\Gamma_i(\infty)(1+g_i/\sigma)$, which reflects the different lifetimes of the localized phonons[8,31,32]. $\Gamma_i(\infty)$ is the linewidth for the bulk crystal. $\Gamma_i(\infty)$ and $g_i$ sets of values were determined from experimental data for bulk crystal and fused silica correspondingly.

For the fused silica sample, the high quality UV-transparent microscope slide was taken (Electron Microscopy Sciences). The Raman measurements were carried out at NCTU with a laboratory built spectroscopic system described elsewhere[33]. Excitation wavelength was 532 nm with laser power of 12 mW at the sample point; 20x objective was used. 8 spectra were averaged with acquisition time of 30s. The measured spectra were reduced with the Boltzmann factor and corrected for the $v^3$ frequency dependence[33].

## Results and discussions

The calculated Raman spectral patterns for different propagation lengths are shown in fig. 1. The bands of the bulk cristobalite broaden asymmetrically upon confinement in accordance with the dispersion functions. The most pronounced size dependence is seen in the low-wavenumber region (LWR). Similar size sensitivity can be found in nanoparticles[17,34] and hydrogen-bonded network of water[13], both experimentally and theoretically.

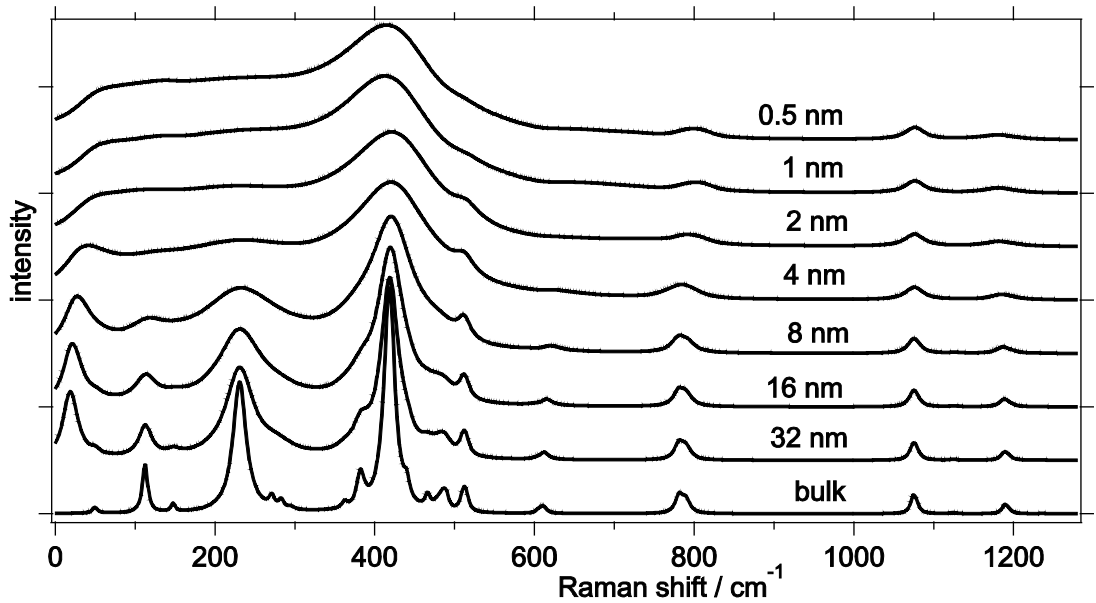

Figure 1. Calculated Raman spectral patterns for different phonon propagation lengths (eq. 2).

For the size below 2 nm, the calculated spectrum converges to the asymptotic pattern (fig. 1). From the Fourier coefficients, $C(q_0,q)$ it can be seen that the q-vectors are effectively spread through the whole Brillouin zone for sub-2-nm confinement (fig. 2).

From comparison of experimental Raman spectra with the calculations (fig. 3), it is clear that the observed spectrum corresponds to the converged spectral pattern, in which phonons from the whole BZ are allowed. For the acoustic phonons considered separately, the calculated band maximum is highly sensitive to the size in the few-nm range (fig. 4). It arrives to asymptote of ~70 cm$^{-1}$ by the sub-1-nm confinement, which corresponds to 2 unit cells. The corresponding cluster of this size is shown in fig. 4 (right).

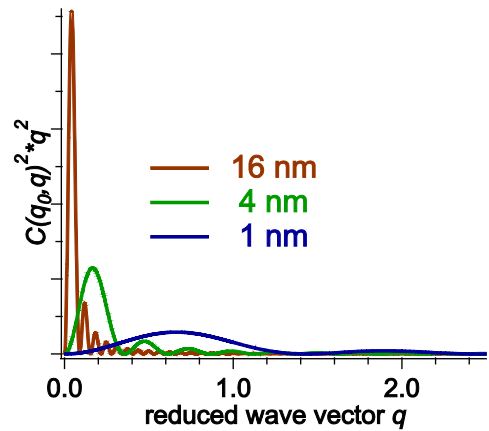

Figure 2. Contribution of $q$-points to the Raman intensity for different phonon propagation lengths

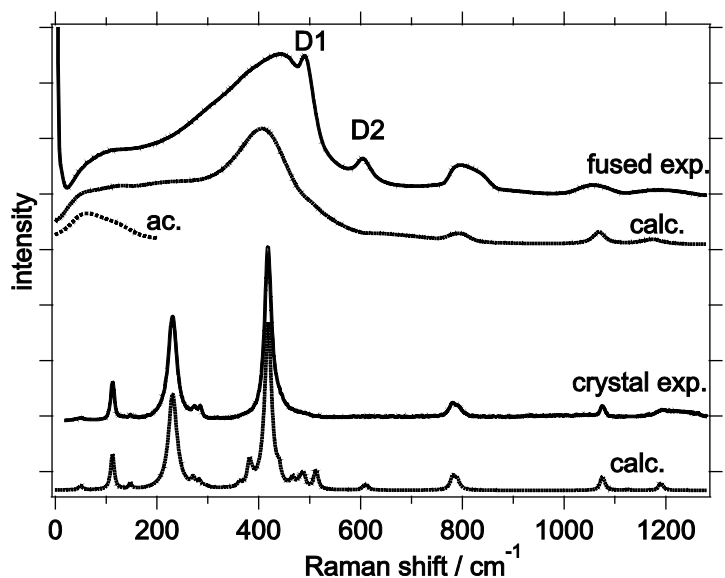

Figure 3. Experimental Raman spectra of fused silica (top, solid line) and cristobalite from [21,35] (bottom, solid line); calculated spectra (eq. 2) are shown with dotted lines. For the latter, the 1 nm phonon propagation was considered. The spectrum from acoustic-only phonons is plotted additionally. Defect bands from 4- and 3-member rings are labeled D1 and D2 respectively [36,37].

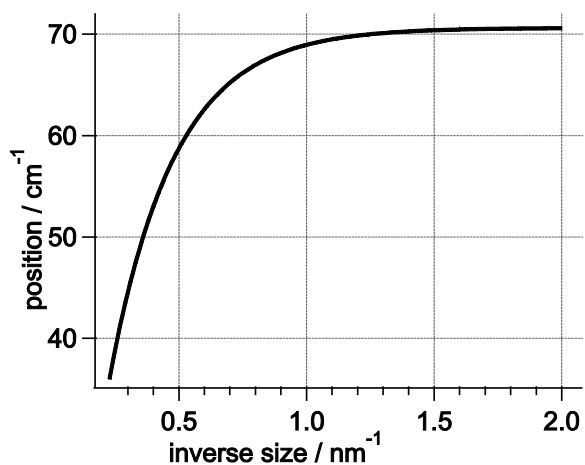
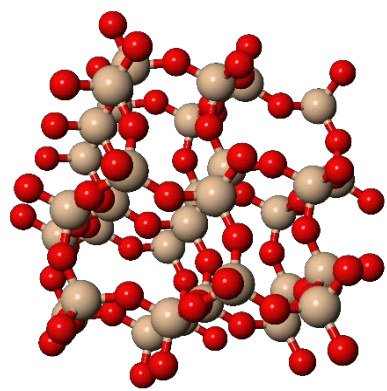

Figure 4. Calculated position of the band maximum for acoustic modes vs. inverse size (left) and the 1-nm cluster with cristobalite-ordered structure (right)

## Conclusions

In conclusion, the Raman spectral pattern of vitreous silica can be described as confined phonons of $SiO_2$ network. Low-wavenumber region of the spectrum has a pronounced dependence on the phonon propagation scale. Upon the confinement, the Raman spectrum converges by the 2 nm size to the spectral pattern, in which $q$-points from the whole Brillouin zone are effectively allowed. Unlike the hydrogen-bonded network of water, the structure of vitreous silica is characterized by high degree of uniformity.

## Acknowledgements

Many thanks to Dr. D.M. Sedlovets for providing the computational facility for this work, Chun-Chieh Yu, Ankit Raj and Prof. Hiro-o Hamaguchi (USILab, NCTU) for help with spectral measurements. This work was supported by the Ministry of Science and Higher Education of the Russian Federation, program no. 075-00475-19-00.